# Derivation of the Schrödinger equation from fundamental principles

Wenzhuo Zhang and Anatoly Svidzinsky*

Department of Physics & Astronomy, Texas A&M University, College Station TX, 77843, USA

*Correspondence: asvid@physics.tamu.edu

**Definition**

Schrödinger's path to the quantum mechanical wave equation was heuristic and guided more by physical intuition than formal deduction. Here we derive the Schrödinger equation for the particle's wave function $\Psi$, assuming that the complex function $\Psi(t, \vec{r})$ has a meaning of the probability amplitude to find the particle at time $t$ at point $\vec{r}$ and the relations $E = \hbar\omega$, $\vec{p} = \hbar\vec{k}$ expressing particle energy and momentum in terms of the frequency and wave vector of the associated probability wave.

## 1. History

### A. Old quantum theory (1900–1925)

The quantum revolution began in 1900 with Max Planck's attempt to explain the spectrum of blackbody radiation. The problem is connected to one of the two "clouds" that, as Lord Kelvin noted, darkened classical physics at the century's end [1]. There was no explanation for the overall shape of the observed emission spectrum of blackbody radiation at that time. Classical calculation based on the equipartition theorem (known as the Rayleigh–Jeans law) predicted an infinite energy emission at high frequencies. Planck first sought to adjust Wien's empirical formula, which worked only at high frequencies, but soon obtained a relation valid at all frequences [2]. He modeled blackbody as a set of harmonic oscillators and assumed that their energy consisted of a definite number of equal portions of size $h\nu$, involving the Planck constant $h$ he had introduced the previous year [3]. This discretization allowed a finite count of possible states in entropy calculation, from which he obtained the blackbody radiation law. In terms of the spectral energy density $u_\nu(\nu, T)$ (energy per unit volume per unit frequency $\nu$ interval) at a given temperature $T$, the Planck's formula reads

$$u_\nu(\nu, T) = \frac{8\pi h\nu^3}{c^3} \frac{1}{e^{h\nu/k_B T} - 1},$$

where $c$ is the speed of light and $k_B$ is the Boltzmann constant.

The match with experiment was accurate at all frequencies. Yet Planck regarded quantization as a mathematical expedient, without committing further to physical interpretation [4]. He unknowingly opened the new quantum era.

Despite Max Planck's reluctance, the idea of energy discreteness soon proved indispensable in explaining the photoelectric effect in which electrons are emitted from metal surfaces that are illuminated by light above a certain threshold frequency, regardless of light intensity. First observed by Heinrich Hertz in 1887 [5], the phenomenon is difficult to explain in the context of classical electromagnetic theory. In 1905, Albert Einstein proposed that light consists of discrete "light quanta" (photons) of frequency $\nu$, each carrying energy $E = h\nu$ and momentum $p = E/c$ [6]. The maximum kinetic energy $K_{max}$ of an ejected electron is given by

$$K_{\max} = h\nu - W,$$

where $W$ is the work function, the minimum energy needed to free an electron from the metal. The linear dependence of $K_{\max}$ on light frequency, not intensity, indicated that each electron absorbs a single quantum of light. Einstein's prediction was later confirmed by Robert Millikan [7] and by Arthur Compton [8], establishing that light has a duality, exhibiting both wave-like and particle-like properties.

In addition to other puzzles, the structure of atoms posed another crisis for classical physics. According to electromagnetic theory, an electron orbiting a positively charged nucleus should radiate energy continuously and collapse into the nucleus, making atomic stability impossible.

In 1913, Niels Bohr proposed a new model to address this paradox [9]. He managed to reconcile the observed hydrogen atom spectrum by introducing quantum rules at the points where classical physics failed. Electrons, he asserted, can occupy only specific stationary states in which they do not radiate. Radiation occurs only when an electron makes a transition between these states, emitting or absorbing a photon of energy $E = h\nu$, equal to the difference between the two energy levels [10].

Bohr's model successfully explained the wavelengths of hydrogen's spectral lines, but its assumptions lacked deeper justification, and the model quickly broke down for more complicated systems such as many-electron atoms and molecules. Even for the hydrogen atom Bohr's model incorrectly predicted nonzero orbital angular momentum of the electron in the ground state. Some of Bohr's original postulates [11] were later shown to be incorrect. Nevertheless, the idea of stationary states and discrete transitions was preserved in later quantum mechanics, where these rules emerged as approximations within a broader framework. Bohr's formulation marked a decisive step in quantizing atomic and molecular structure.

While foregoing developments applied discreteness to various physical models, Louis de Broglie pursued the converse path—seeking to restore continuity. In his 1924 doctoral thesis, he proposed that every material particle is accompanied by a matter wave [12]. He associated angular frequency $\omega$ and wave vector $\vec{k}$ of the plane matter wave $e^{-i\omega t + i\vec{k}\cdot\vec{r}}$ with the particle's energy $E$ and momentum $\vec{p}$, introducing

$$E = \hbar\omega, \qquad \vec{p} = \hbar\vec{k}, \qquad (1)$$

in analogy to Einstein's photon (here $\hbar = h/2\pi$).

De Broglie further argued that his hypothesis could account for some previously discussed quantum phenomena [12]. For instance, a stable Bohr orbit, he suggested, corresponds to a situation where the associated electron matter wave remains in phase with itself after one complete revolution—a standing wave condition. This idea gave Bohr's quantization rule a physical interpretation based on wave coherence and later served as the foundation for the quantization condition of Bohr and Sommerfeld.

Although de Broglie did not yet provide a governing wave equation from which these matter waves could be derived, his theory established a profound unification between radiation and matter and directly inspired Erwin Schrödinger's formulation of wave mechanics.

De Broglie's doctoral thesis was not accepted right away: Paul Langevin, de Broglie's advisor, regarded his work as unconventional and sought external judgment before approval. He shared the thesis with his friend Einstein, who immediately realized that this was a good thing and cited de Broglie's ideas in his own work, thereby bringing them to wider attention. Schrödinger learned of de Broglie's hypothesis through Einstein's paper and, after some effort, obtained a copy of the thesis around November 1, 1925, and in the sequel laid down mathematical foundation of wave mechanics. Schrödinger's first paper on wave mechanics was submitted two months after he received de Broglie's thesis, on January 27, 1926. The subsequent papers were submitted at a rate of about one per month, on February 23, March 18, May 10, June 21.

### B. Schrödinger's effort

Erwin Schrödinger was somewhat marginal in quantum science before 1925 [13]. Though not previously central to the field, he published the Schrödinger equation in early 1926, and thereby established wave mechanics [14]. His background and intellectual formation, however, gave him decisive advantages that made such a discovery possible.

As a student at the University of Vienna, Schrödinger inherited the academic tradition of Ludwig Boltzmann. Boltzmann promoted the atomic theory with his indispensable contributions to statistical mechanics, at a time when many physicists still doubted the existence of atoms [15]. Schrödinger absorbed this tradition, pursuing problems such as quantum gases from the perspective of statistical mechanics. His study of works by M. Planck, A. Einstein, and S. N. Bose convinced him that quantum gases should be treated as holistic systems rather than mere collections of particles [13]. This statistical perspective later prepared him to apply de Broglie's wave vision to the hydrogen spectrum.

Schrödinger encountered similar occasions in 1922 in studying Hermann Weyl's gauge theory [14], which revealed the role of quantum conditions. There, he accumulated insights resurfaced later accepting complex wavefunctions as natural mathematical tools [16].

Schrödinger received his major inspiration from de Broglie, as he later acknowledged. Yet de Broglie's ideas were not widely accepted at the time. De Broglie disagreed with Bohr and Bohr's supporters on several issues, placing his ideas outside mainstream physics. Apart from this, de Broglie's proposal on associating waves with material particles was sometimes viewed as an imitation of earlier attempts to treat gravitational waves, and thus less impressive [14]. Only a few physicists, most notably Einstein, saw its potential. Einstein cited de Broglie approvingly and helped disseminate his ideas. Having previously translated and generalized Bose's work on statistical mechanics [17], Einstein indirectly shaped Schrödinger's thinking. Einstein's influence encouraged Schrödinger to unite de Broglie's wave concept with Einstein's long-lasting preference for differential equations, leading to the Schrödinger equation.

At the end of 1925, Schrödinger gave a colloquium about de Broglie's ideas in Zurich at Peter Debye's invitation. Debye's skeptical remark—that de Broglie's theory required a proper wave equation—provided the final impetus [18]. Schrödinger was prompted by de Broglie's plane wave related to a particle in uniform motion, and conjectured a wave equation for an electron in a Coulomb potential. His first attempt yielded the relativistic Klein–Gordon equation. Failed to incorporate electron spin appropriately, the equation produced incorrect spectrum for hydrogen atom [13].

What seems to have been an inspired afterthought, Schrödinger turned from an unsuccessful relativistic attempt to explore a non-relativistic formulation. Between January and June 1926, he published four papers titled *Quantisierung als Eigenwertproblem*, which together trace the gradual development of his approach to the wave equation.

In the first paper, following Debye's suggestion to use the variational principle, Schrödinger "derived" the equation for stationary states. Although he did not yet grasp the meaning of the assumed function $\Psi$, the equation reproduced the accepted hydrogen emission spectrum [19]. In retrospect, this success was largely a serendipity. Schrödinger took as a primary postulate the classical Hamilton–Jacobi equation for a nonrelativistic particle constrained by a time-independent potential. The following derivation of the action for $\Psi$ was mathematically incorrect, in addition to the assumption that the wave function $\Psi$ is a real-valued function. Perhaps Schrödinger knew the equation for $\Psi$ which yields the right hydrogen spectrum and tried to find a way to reach it (reverse engineering). He soon recognized that the action introduced could not be properly justified. Even after correcting it, he offered insufficient explanation for why this particular form of action should prevail over others.

In the second paper, Schrödinger embraced the wave picture and justified his "undulatory" mechanics through Hamilton's optical–mechanical analogy: as geometrical optics fails for short wavelengths or strong curvatures, so does the classical mechanics fail in the atomic regime. Hence, a wave mechanics might succeed as wave optics had once replaced geometrical optics. From this standpoint, Schrödinger constructed equation directly from the classical wave equation [20].

The result again described stationary states. The full time-dependent form of the equation was presented only in the Schrödinger's fourth paper, though some hints on the temporal behavior had appeared earlier. Accepting complex functions, and motivated by the structure of the classical Hamilton-Jacobi equation, he introduced a term with a first-order time derivative [21]. The equation, though resembling a diffusion equation rather than a wave equation, governs the temporal evolution of the matter wave.

To summarize, Schrödinger's path in obtaining the main equation of quantum theory was heuristic and guided more by physical intuition than formal deduction. Schrödinger proposed the equation without rigorous mathematical derivation from the underlying fundamental principles which were unknown at that time. Result of Schrödinger's efforts was an equation for a complex-valued matter wave function $\Psi(t, \vec{r})$, whose evolution follows a linear differential equation linking the function time derivative to the system's total energy:

$$i\hbar \frac{\partial}{\partial t} \Psi(t, \vec{r}) = -\frac{\hbar^2}{2M} \nabla^2 \Psi(t, \vec{r}) + V(t, \vec{r}) \Psi(t, \vec{r}).$$

Here, the first term on the right-hand side represents kinetic energy of a nonrelativistic particle with mass $M$, and the second term is the particle's potential energy. For a system in a stationary state, the left-hand side reduces to $E\Psi(t,\vec{r})$, yielding the familiar energy eigenfunctions that define the stationary solutions of the Schrödinger equation.

### C. Later developments

After his second paper, Schrödinger demonstrated that wave mechanics was mathematically equivalent to the matrix mechanics developed slightly earlier by Werner Heisenberg, Max Born, and Pascual Jordan [22].

In the same year, Born introduced a new interpretation: the squared modulus of the wave amplitude, $|\Psi(t,\vec{r})|^2$, represents the probability density of finding particle at position $\vec{r}$ and time $t$ [23]. Since wave functions superpose, this view explained interference naturally and led to Born rule, which underlies all empirically successful formulations of quantum mechanics.

Soon afterward, Paul A. M. Dirac formulated a relativistic wave equation for the electron, successfully incorporating its spin and predicting the existence of antimatter [24]. He also established the first consistent quantum description of the electromagnetic field [25], laying the foundations of quantum field theory.

These achievements, made within the short period from 1925 to 1928, completed the foundational formation of quantum mechanics [26]. Although philosophical debates continued, quantum theory agreed with experiments to extraordinary precision and became one of the most successful frameworks in physics.

Among these advances, wave mechanics holds a special place. It conveys the intuition of matter "flowing" as a continuous wave and relies on differential equations to describe its evolution. Schrödinger's equation lies at the heart of every modern quantum physics course, yet it is introduced and explained to students without proper derivation. The latter should begin from fundamental principles, rather than follow Schrödinger's heuristic path. In the following we present a tutorial derivation of the Schrödinger equation from the fundamental principles which underline the probabilistic nature of wave mechanics.

## 2. Derivation of the Schrödinger equation

Quantum mechanics is based on the concept that fundamental laws of physics, manifesting at the microscopic level, are probabilistic in nature rather than deterministic. Namely, in classical mechanics particles move along unique trajectories determined by the initial conditions. In quantum (wave) mechanics particles do not follow classical trajectories and particle motion is described by the evolution of the associated probability wave. The latter is characterized by a complex wave function $\Psi(t,\vec{r})$ which has a meaning of the probability amplitude for detecting the particle at position $\vec{r}$ at time $t$ (Born rule). That is probability to find the particle in the spatial volume element $d^3r$ at time $t$ is equal to $d^3r|\Psi|^2$. $\Psi(t,\vec{r})$ carries information about probability distribution of particle position, momentum, energy, etc.

To be specific, we consider a nonrelativistic particle of mass $M$ which moves in an external potential $V(t,\vec{r})$. In classical physics the total mechanical energy of the particle $E$ is given by

$$E(t,\vec{r}) = E_{kin}(t,\vec{r}) + V(t,\vec{r}), \quad (2)$$

where in classical mechanics

$$E_{kin} = \frac{p^2}{2M}$$

is the particle's kinetic energy and $\vec{p}$ is momentum.

In the quantum mechanical description, we still require that relation (2) holds locally (at any $t$ and $\vec{r}$). That is, if the particle is located in the volume element $d^3r$ at time $t$, the total mechanical energy of the particle is equal to the kinetic plus potential energy. But now $E(t, \vec{r})$ and $E_{kin}(t, \vec{r})$ are determined by $\Psi$. To obtain this dependence we use the Planck–de Broglie relations (1) which hold for states with certain frequency $\Psi(\vec{r})e^{-i\omega t}$ and certain momentum $\Psi(t)e^{i\vec{k}\cdot\vec{r}}$ respectively.

Expanding $\Psi(t, \vec{r})$ in Fourier series in $\vec{k}$ we have

$$\Psi(t, \vec{r}) = \frac{1}{(2\pi)^{3/2}} \sum_{\vec{k}} \Psi_{\vec{k}}(t) e^{i\vec{k}\cdot\vec{r}}, \quad (3)$$

where $|\Psi_{\vec{k}}|^2$ is the probability that particle has momentum $\vec{p} = \hbar\vec{k}$. Using Eq. (3), the average momentum of the particle

$$\langle \vec{p} \rangle = \sum_{\vec{k}} \hbar\vec{k} |\Psi_{\vec{k}}|^2$$

can be written as

$$\langle \vec{p} \rangle = -i\hbar \int d^3r \frac{1}{2}(\Psi^*\nabla\Psi - \Psi\nabla\Psi^*) = \hbar \int d^3r |\Psi|^2 \nabla S, \quad (4)$$

where $S$ is the phase of the wave function, $\Psi = |\Psi|e^{iS}$. Similarly, the average kinetic energy of the particle

$$\langle E_{kin} \rangle = \sum_{\vec{k}} \frac{\hbar^2 k^2}{2M} |\Psi_{\vec{k}}|^2$$

can be written as

$$\langle E_{kin} \rangle = -\frac{\hbar^2}{2M} \int d^3r \frac{1}{2}(\Psi^*\nabla^2\Psi + \Psi\nabla^2\Psi^*) = \frac{\hbar^2}{2M} \int d^3r |\Psi|^2 \left((\nabla S)^2 - \frac{\nabla^2 |\Psi|}{|\Psi|}\right), \quad (5)$$

where we used equation

$$\int e^{i(\vec{k}-\vec{k}')\vec{r}} d^3r = (2\pi)^3 \delta(\vec{k} - \vec{k}').$$

Since $d^3r|\Psi|^2$ is the probability to find particle in the volume element $d^3r$, Eqs. (4) and (5) yield that the local values of the particle momentum and kinetic energy are given by

$$\vec{p} = \hbar\nabla S, \quad (6)$$

and

$$E_{kin} = \frac{\hbar^2}{2M}\left((\nabla S)^2 - \frac{\nabla^2 |\Psi|}{|\Psi|}\right) = \frac{p^2}{2M} - \frac{\hbar^2}{2M}\frac{\nabla^2 |\Psi|}{|\Psi|} \quad (7)$$

respectively. They are determined by $\Psi$. Please note that $E_{kin}$ differs from the classical expression $p^2/2M$ by the energy of particle confinement (the last term on the right-hand-side of Eq. (7)). The latter is also known as the Bohm quantum potential energy introduced by David Bohm in 1952. It is nonlocal; can be both positive and negative; and it does not change if $|\Psi|$ is multiplied by a constant, as this term is also present in the denominator. The mean value of the quantum potential is proportional to the Fisher information (*FI*) about the particle position. The latter property follows from the identity

$$FI = \int |\Psi|^2 (\nabla ln |\Psi|^2)^2 d^3r = -4\int |\Psi|^2 \frac{\nabla^2 |\Psi|}{|\Psi|} d^3r.$$

To find expression for the local value of the total particle energy, we expand $\Psi(t,\vec{r})$ in terms of the energy eigenstates $\Psi_\omega(\vec{r})$, which for simplicity we assume to be discrete

$$\Psi(t,\vec{r}) = \sum_\omega C_\omega \Psi_\omega(\vec{r}) e^{-i\omega t}, \quad (8)$$

where $|C_\omega|^2$ is the probability that particle has energy $\hbar\omega$. Using Eq. (8), the average total energy of the particle

$$\langle E \rangle = \sum_\omega \hbar\omega |C_\omega|^2$$

can be written as

$$\langle E \rangle = i\hbar \int d^3r \frac{1}{2}\left(\Psi^* \frac{\partial \Psi}{\partial t} - \Psi \frac{\partial \Psi^*}{\partial t}\right) = -\hbar \int d^3r |\Psi|^2 \frac{\partial S}{\partial t}, \quad (9)$$

where we used the orthogonality relation

$$\int d^3r \Psi^*_{\omega'}(\vec{r}) \Psi_\omega(\vec{r}) = \delta_{\omega'\omega}.$$

Thus, the local value of the total particle energy is given by

$$E = -\hbar \frac{\partial S}{\partial t}. \quad (10)$$

Plug Eqs. (7) and (10) in Eq. (2) yields the quantum mechanical analog of the classical Hamilton-Jacobi equation

$$-\hbar \frac{\partial S}{\partial t} = \frac{\hbar^2}{2M}\left((\nabla S)^2 - \frac{\nabla^2 |\Psi|}{|\Psi|}\right) + V. \quad (11)$$

One should note that in quantum mechanics the local value of kinetic energy (7) can be negative, which corresponds to tunneling into the classically forbidden regions.

Since probability is conserved, $\Psi(t,\vec{r})$ must also obey the continuity equation

$$\frac{\partial |\Psi|^2}{\partial t} + \nabla \cdot \vec{j} = 0, \quad (12)$$

where the probability current density is given by

$$\vec{j} = |\Psi|^2 \vec{u}, \quad (13)$$

and $\vec{u}$ is the local particle velocity which, with the help of Eq. (6), reads

$$\vec{u} = \frac{\vec{p}}{M} = \frac{\hbar \nabla S}{M}. \quad (14)$$

Taking into account that $\Psi = |\Psi|e^{iS}$ and

$$\nabla S = \frac{i}{2|\Psi|^2}(\Psi \nabla \Psi^* - \Psi^* \nabla \Psi), \quad (15)$$

Eqs. (11) and (12) can be written as a single equation for the complex function $\Psi$

$$i\hbar \frac{\partial \Psi}{\partial t} = -\frac{\hbar^2}{2M}\nabla^2 \Psi + V(t, \vec{r})\Psi. \quad (16)$$

Equation (16) is known as the Schrödinger equation for a single nonrelativistic particle moving in potential $V(t, \vec{r})$. Our derivation of the Schrödinger equation implies that the particle is characterized by the local velocity $\vec{u}(t, \vec{r})$ and probability density $|\Psi(t, \vec{r})|^2$. This is analogous to the hydrodynamic description of fluid motion and leads to hydrodynamic interpretation of quantum mechanics. However, one should emphasize that the local velocity $\vec{u}(t, \vec{r})$ and the local momentum $\vec{p}(t, \vec{r}) = M\vec{u}(t, \vec{r})$ are not obtained as results of measurements on the state of the particle. If the particle is detected in the volume element $d^3r$, the wave function collapses into a state localized in this volume, and momentum distribution is now determined by the collapsed $\Psi$. In an ideal momentum measurement, the detector is designed such that $\Psi$ collapses into the momentum eigenstate $e^{i\vec{k}\cdot\vec{r}}$, for which the particle position is equally probable for any $\vec{r}$; that is, the position is uncertain.

One should note that the energy of particle confinement $-\frac{\hbar^2}{2M}\frac{\nabla^2|\Psi|}{|\Psi|}$ is what makes wave mechanics different from the classical mechanics. For macroscopic objects or fast-moving particles, this energy can be neglected, and Eq. (11) reduces to the classical Hamilton-Jacobi equation

$$-\hbar\frac{\partial S}{\partial t} = \frac{\hbar^2}{2M}(\nabla S)^2 + V$$

for the particle action $\hbar S$, which, together with the continuity equation

$$\frac{\partial |\Psi|^2}{\partial t} + \nabla(|\Psi|^2 \vec{u}) = 0, \quad \vec{u} = \frac{\hbar \nabla S}{M}$$

has a $\delta$-function solution

$$|\Psi|^2 = \delta(\vec{r} - \vec{r}(t)), \quad \frac{d}{dt}\vec{r}(t) = \vec{u},$$

describing motion of a localized particle along the classical trajectory $\vec{r}(t)$ with velocity $\vec{u}$.

One can also obtain Schrödinger Eq. (16) if instead of Eq. (2) we assume that quantum mechanics obey the superposition principle. Indeed, plug Eq. (15) into Eqs. (13) and (14) yields

$$\vec{j} = \frac{i\hbar}{2M}(\Psi \nabla \Psi^* - \Psi^* \nabla \Psi). \quad (17)$$

Substitute Eq. (17) into the continuity Eq. (12) gives

$$\frac{\partial}{\partial t}(\Psi\Psi^*) + \frac{i\hbar}{2M}(\Psi\nabla^2\Psi^* - \Psi^*\nabla^2\Psi) = 0.$$

Multiplying both sides of this equation by $i\hbar/\Psi\Psi^*$, we obtain

$$\frac{i\hbar}{\Psi}\left(\frac{\partial \Psi}{\partial t} - \frac{i\hbar}{2M}\nabla^2\Psi\right) = -\frac{i\hbar}{\Psi^*}\left(\frac{\partial \Psi^*}{\partial t} + \frac{i\hbar}{2M}\nabla^2\Psi^*\right) = V, \qquad (18)$$

where $V$ is a function which we need to find. Equation (18) shows that the combination on the left-hand-side is equal to its complex conjugate. Therefore, $V$ is a real function ($V = V^*$). Equation (18) can be written as

$$i\hbar\frac{\partial \Psi}{\partial t} + \frac{\hbar^2}{2M}\nabla^2\Psi = V\Psi. \qquad (19)$$

To obtain Schrödinger equation we can postulate that quantum mechanics obeys the superposition principle, that is, equation for $\Psi$ must be linear. Then, according to Eq. (19), $V$ should be independent of $\Psi$ and $\Psi^*$, and function $V(t,\vec{r})$ must be identified with the particle potential energy. This can be seen by considering a particular solution in the limit when particle moves very fast ($k$ is large). In this limit, $V(t,\vec{r})$ can be treated as a slowly varying function of $t$ and $\vec{r}$, that is $V(t,\vec{r})$ is approximately constant. Then Eq. (19) has a plain-wave solution

$$\Psi = e^{-i\omega t + i\vec{k}\cdot\vec{r}}, \qquad (20)$$

where $\omega$ is the wave angular frequency. Plug Eq. (20) into Eq. (19) yields

$$V = \hbar\omega - \frac{(\hbar k)^2}{2M}. \qquad (21)$$

That is $V$ is the difference between the total particle energy $\hbar\omega$ and its kinetic energy $p^2/2M$. Thus, $V$ is the particle potential energy.

Schrödinger equation (16) can be written as

$$i\hbar\frac{\partial \Psi}{\partial t} = \hat{H}\Psi,$$

where $\hat{H}$ is Hermitian operator

$$\hat{H} = -\frac{\hbar^2\nabla^2}{2M} + V(t,\vec{r})$$

called Hamiltonian. The hermicity of Hamiltonian implies that evolution of the wave function is unitary and energy eigenvalues are real. Hermitian operators are fundamental in quantum mechanics, representing physical observables like position, momentum, and energy.

In 1926, Erwin Madelung [27] showed that if one writes the wave function in the form $\Psi = |\Psi|e^{iS}$ the Schrodinger equation (16) yields that $S$ is governed by the quantum Hamilton-Jacobi equation (11), while $|\Psi|^2$ obeys the hydrodynamic continuity equation (12). Since that time the Madelung equations (11) and

(12) have provided the basis for numerous classical interpretations of quantum mechanics, including the hydrodynamic interpretation proposed by Madelung himself [27—36], the theory of stochastic mechanics due to Nelson and others [37—60], and the hidden-variable and double-solution theories of Bohm and de Broglie, respectively [61—64].

Several alternative ways to obtain Schrödinger equation have been discussed in the literature. For example, it can be obtained from a mathematical identity by a slight generalization of the formulation of classical statistical mechanics based on the Hamilton–Jacobi equation [65]. Schrödinger equation can be also obtained from the theory of Markoff processes [66]; from the Liouville equation [67]; and from stochastic electrodynamics [68]. Schrödinger equation for free particle can be obtained from the quantum field route by using Lorentz boost operation [69], as well as quantum optical and classical Maxwell equations routes [70].

One should also mention the path integral formulation of quantum mechanics of Feynman which emerged from Feynman's discovery that Schrödinger equation follows from a proper choice of a propagator involving classical action of the particle [71,72].

## 3. Conclusions and Prospects

It is believed by many that the Schrödinger equation is one of the postulates of quantum mechanics and it should be taken for granted. In this entry, we convey the message that equations of physical theory can be obtained from the underlying fundamental principles. The present tutorial derivation of the Schrödinger equation from the de Broglie's relation between particles' momentum and wave vector of the associated matter wave, together with the Planck energy–frequency relation and Born rule, is an example.

Historically, many physical theories were not derived from the fundamental principles at first but rather their development was guided by experiments. Only later were the underlying physical principles understood. The theory of electromagnetism is a good illustration. Maxwell's equations that, together with the Lorentz force law, form the foundation of classical electromagnetism were a comprehensive summary of experimental results and rules. Only later, after special relativity was developed, was it realized that Maxwell's equations and the Lorentz force can be obtained in a unique way by postulating that electromagnetic field is a four-vector field in Minkowski spacetime produced by conserved electric charges and currents. The latter implies that the theory must be gauge- and Lorentz-invariant, which, with the help of the least action principle, leads to Maxwell's equations and the Lorentz force.

Mathematical self-consistency of the theory requires the adoption of the "symmetry protocol" in construction field equations. This approach allows us to obtain equations of the theory in a unique way from the underlying fundamental principles. For example, in the case of electromagnetism, the "symmetry protocol" requires that all terms in the action must possess the same symmetries (be gauge- and Lorentz-invariant).

Similarly, with the help of the "symmetry protocol", Einstein's general relativity uniquely follows from a single postulate that spacetime geometry is gravitational field. Mathematically, this assumption implies that the action must be invariant under general coordinate transformations. If we choose the basic principles differently, we will obtain a different theory of gravity. For example, the assumption that the universe has a fixed Euclidean background geometry and gravity is a vector field in the four-dimensional Euclidean space, yields an alternative vector theory of gravity that predicts no spacetime singularities such as black holes, but is equivalent to general relativity in the post-Newtonian limit and passes all other gravitational tests [73]. In contrast to general relativity, vector gravity explains the nature of dark energy as the energy of gravitational field induced by the universe expansion and, with no free parameters, correctly predicts the measured value of the cosmological constant [73]. Vector gravity also explains charged elementary particles as bound states of fundamental fields held together by gravity, and predicts particle masses (e.g. electron and muon) in perfect agreement with experiment without free parameters [74]. In addition, it predicts the vector polarization of gravitational waves, which was confirmed by LIGO/Virgo gravitational wave detection [75]. That is, modification of the underlying principles alters predictions of the theory, which can be tested experimentally. In the case of gravity, experiments indicate that geometry of the universe is not dynamical, as postulated by Einstein, but rather universe has a fixed background geometry, as assumed in vector gravity.

Modern physics greatly benefits from the symmetry approach. For example, the Standard Model of particle physics is based on postulates of symmetry (Lorentz and gauge invariance, specifically the gauge group $SU(3) \times SU(2) \times U(1)$), which turns out to be amazingly powerful and accurately accounts for many physical observations. The Model classifies all known elementary particles and describes three of the four fundamental forces in nature: electromagnetic, weak, and strong interactions. One should note, however, that the Standard Model provides only a phenomenological description of elementary particles—its symmetries and basis states were largely selected based on the enormous experimental work that engaged the entire global community over the course of over a century. Many aspects of this model (multiplicity of families of elementary particles, origin of particle charges, etc.) remain unexplained. Gravity is not part of the Standard Model because Einstein's general relativity cannot explain what elementary particles are and is incompatible with quantum theory. In contrast, vector gravity fits well into particle physics, and is the way to go [74].

Human curiosity drives the quest to find the most fundamental building blocks of nature and understand how the universe was created. Perhaps we must figure this out from pure thought, since no experiment can reproduce the extreme conditions of the Big Bang. The emergence and comprehension of physical theories from the perspective of fundamental principles will drive this quest in years to come.

**Acknowledgments:** We are grateful to Prof. Marlan Scully for valuable discussion.

**Funding:** This research was funded by U.S. Department of Energy (DESC-0023103, FWP-ERW7011, DE-SC0024882); Welch Foundation (A-1261); National Science Foundation (PHY-2013771); Air Force Office of Scientific Research (FA9550-20-1-0366). W.Z. is supported by the Herman F. Heep and Minnie Belle Heep Texas A&M University Endowed Fund held and administered by the Texas A&M Foundation.